\documentclass[aps,prl,twocolumn,groupedaddress]{revtex4}
\usepackage{graphicx}
\begin{document}

\title{Heat capacity of liquids: an approach from the solid phase}
\author{Kostya Trachenko}
\address{Department of Earth Sciences, University of Cambridge,
Downing Street, Cambridge, CB2~3EQ, UK}
\begin{abstract}
We calculate the energy and heat capacity of a liquid on the basis of its elastic properties and vibrational states. The
experimental decrease of liquid heat capacity with temperature is attributed to the increasing loss of two transverse modes with
frequency $\omega<1/\tau$, where $\tau$ is liquid relaxation time. In a simple model, liquid heat capacity is related to viscosity and is compared with the experimental data of mercury. We also calculate the vibrational energy of a quantum liquid, and show that transverse phonons can not be excited in the low-temperature limit. Finally, we discuss the implications of the proposed approach to liquids for the problem of glass transition.
\end{abstract}


\maketitle

\section{Introduction}

Heat capacity is one of the most important physical properties of a system, because it holds information about its degrees of freedom. Heat capacity is well understood in gases and solids. Liquids, on the other hand, remain an exception, with the result that their heat capacity is often barely discussed in classic statistical physics text books \cite{landau} or texts dedicated to liquids \cite{frenkel,ziman,yip,march,zwanzig,hansen}. The general difficulty in calculating the energy and other thermodynamic quantities of a liquid is that, unlike a gas, it has strong interactions, but at the same time does not have small atomic vibrations as a solid \cite{landau}. Because the inter-molecular interactions in a liquid are strong, the properties of this interaction, and, therefore, properties of the liquid, are strongly system-dependent \cite{landau}. Consequently, it is argued that no general recipe for calculating the energy of a liquid can exist \cite{landau}.

Experimental data from many elemental liquids show that their constant-volume heat capacity, $C_v$, {\it decreases} with temperature, from about $3N$ at the melting point to $(2-2.5)N$ at high temperature \cite{grimvall,wallace}. Here, $N$ is the number of particles and $k_{\rm B}=1$. The decrease of $C_v$ is also seen in many complex liquids \cite{dexter}. This behaviour is not understood in a consistent framework similar to those that exist for solid and gas phases.

Most calculations of liquid energy, $E$, approach a liquid from the gas phase, and calculate the potential energy in addition to the gas kinetic energy, giving $E=K+U$, where $K$ is the kinetic energy of an ideal gas ($K=3NT/2$) and $U$ is the potential energy of interatomic interactions \cite{landau,frenkel,ziman,march}. The simplest treatment assumes the case of a dilute system with only pair interactions. Assuming that the interactions are weak (except at short separations), the high-temperature expansion gives the system described by the Van Der Waals equation \cite{landau}. While this approach can describe a dense gas or a liquid close to the critical point, it is not adequate for real liquids. For example, $C_v$ of the Van Der Waals is equal to that of an ideal gas \cite{landau}, in contrast with the experimental results. The improved calculations of $U$ employ higher-order (three- and four-) particle correlations \cite{march}. The results, however, are not straightforward to use for numerical estimates \cite{march} and require the knowledge of correlation functions as well as the properties of interatomic interactions.

A distinctly different, and less common, approach to discuss the energy and heat capacity of a liquid is to include the strong interactions from the outset, by approaching a liquid from the solid phase. The temperature-dependent term of the energy of a solid is given by the phonon energy. In an isotropic solid (glass) all vibrations can be represented by one longitudinal and two transverse waves. In classical case, this gives the heat capacity of $C_v=3N$, the Dulong-Petit result. The discussion then proceeds to establish how this result needs to be modified on transition from the solid to the liquid phase. Therefore, this approach accounts for the strong interactions in a liquid from the outset. In this sense, it is opposite to the approach that starts with the energy of an ideal gas and introduces interactions as a correction.

In this approach, Brillouin modified the solid-like result, $C_v=3N$, by assuming that a liquid does not support transverse
waves, and encountered an interesting contradiction \cite{bril}. The total energy of a solid is $E=(NT/2+NT/2)+2(NT/2+NT/2)=3NT$, where the first and the second term give the energy of longitudinal and transverse
vibrations, respectively, and $NT/2$ is the mean potential or kinetic energy. If a liquid loses two transverse vibrations,
their potential terms vanish, but kinetic terms remain, and the liquid energy becomes $E=(NT/2+NT/2)+2(NT/2)=2NT$, giving $C_v=2N$ \cite{bril}. This contradicted the experimental result that liquid heat capacity at the melting point is about the same as that of crystals, $C_v=3N$. One way to resolve the contradiction is to assume that a liquid consists of small crystalline domains that support transverse waves in some directions, giving $C_v=3N$, but not in others, enabling a liquid to flow \cite{bril}.

$C_v$ was also studied in molecular dynamics simulations, with liquid Al as a case study \cite{md}. Similar to the experiment, the decrease of heat capacity with temperature was observed, and was interpreted as the progressive loss of liquid shear resistance \cite{md}.

The idea that a liquid loses two transverse vibrational modes was also used by Landau to calculate the energy of phonon excitations in a quantum liquid at very low temperature. This gives the result that the temperature-dependent energy term of a liquid is three times smaller than that in a harmonic solid, reflecting the fact that only one longitudinal mode is preserved \cite{landau}.

More recently, liquid heat capacity has been discussed on the basis of alternative mechanisms, including the consideration of liquid potential energy landscape with intervalley motions \cite{wallace}.

In this paper, we propose that energy and heat capacity of liquids can be understood on the basis of their solid-like elastic properties. In particular, we discuss how the idea of relaxation time $\tau$ can be used to describe liquid vibrational states by approaching them from the solid phase. The experimentally observed decrease of liquid heat capacity with temperature is attributed to the increasing loss of two transverse waves with frequency $\omega<1/\tau$. In a simple model, we relate liquid heat capacity and viscosity, and compare this relation with the experimental data of mercury. In addition, we calculate the vibrational energy of a quantum liquid, and show that transverse phonons can not be excited in the low-temperature limit. Finally, we discuss the implications of the proposed approach to liquids for the problem of glass transition.

\section{Experimental evidence for propagating modes in liquids}

The existence of dispersion relations and the ability to support propagating shear modes has been traditionally assigned to solids. Relatively recently, experiments have shown that these features also exist in liquids, with the evidence coming primarily from inelastic X-ray, neutron and Brillouin scattering experiments. In this section, we briefly recall some of these results.

Early experiments detected the presence of propagating modes in liquids, in the form of ``dispersion curves'' at high frequency, at the temperature around melting \cite{copley}. Later, the measurements were extended to temperatures considerably above the melting point, and confirmed the existence of collective excitations in many liquids (see, for example, Refs. \cite{pilgrim,burkel,rec-review,grim}). As reviewed recently, it is now well established that dynamics in liquids shows solid-like character, in that liquids can sustain high-frequency propagating modes down to wavelengths on the atomic scale, with solid-like dispersion relations \cite{rec-review}.

The solid-like ability of a liquid to support high-frequency shear waves considerably above the melting point was directly observed some time ago \cite{grim}. In addition, the ability to support shear waves has been inferred from the widely observed ``positive dispersion'', the increase of high-frequency sound velocity over that expected on the basis of hydrodynamics \cite{pilgrim,burkel,rec-review}. The general explanation of this effect is that at high frequency, a liquid responds elastically to shear stress, developing an ability to support solid-like shear vibrations, in addition to longitudinal ones. As a result, the propagating speed starts approaching that in a solid, i.e. increases. Initially, these observations were made close to the melting point, and were attributed to the proximity of solid phase. However, later experiments found positive dispersion at temperatures considerably above the melting point, confirming the general idea that liquids support solid-like shear waves at high frequencies that extend down to microscopic wavelengths \cite{rec-review}.

\section{View of a liquid from the solid phase: relaxation time}

Relaxation time of a liquid was phenomenologically introduced by Maxwell in the viscoelastic picture of flow \cite{max}. This picture is based on the assumption that deformation of a liquid in response to stress can be interpolated as the sum of elastic and viscous terms, giving $\frac{{\rm d}v_x}{{\rm d}y}=\frac{1}{G}\frac{{\rm d} P_{xy}}{{\rm d}t}+\frac{1}{\eta}P_{xy}$, where $P_{xy}$, $G$ and $\eta$ are shear stress, shear modulus and viscosity, respectively, and the first and the second terms represent elastic and viscous response, respectively. When external motion stops, $v_x=0$, stress relaxes as $P=P_0\exp(-t/\tau)$, where $\tau=\eta/G$ is the Maxwell shear relaxation time.

Frenkel offered microscopic interpretation of $\tau$ as the time between elementary rearrangement processes in a liquid \cite{frenkel}. He started the discussion with the contradiction of essentially the same origin as encountered by Brillouin: on one hand, in order to explain the solid-like value of heat capacity in liquids at the melting point, $C_v=3N$, thermal motion in liquids should be considered to be solid-like, i.e. vibration around fixed positions. On the other hand, this picture contradicts fluidity of liquids. Frenkel reconciled the contradiction by introducing $\tau$ as the time between consecutive atomic jumps from one equilibrium position to another in a liquid. If $\tau$ is large compared with the period of atomic vibrations, a liquid is characterized by vibrational states as a solid, giving $C_v=3N$. If, at the same time, $\tau$ is small compared with time $t$ during which an external force acts on a liquid, usual liquid flow takes place. If, on the hand, $\tau$ is large compared with $t$, a liquid responds to an external force elastically as a solid. Hence the elastic and viscous regime correspond to $\omega\tau\gg 1$ and $\omega\tau\ll 1$, respectively, where $\omega$ is the frequency of external force. This can be further illustrated by considering an external force $F=A\exp(i\omega t)$, for which Maxwell's interpolation gives $\frac{{\rm d}v_x}{{\rm d}y}=\frac{1}{\eta}(1+i\omega\tau)F$, hence $\omega\tau\gg 1$ and $\omega\tau\ll 1$ corresponds to elastic and viscous response, respectively.

In this picture, Frenkel arrived at an important conclusion about how the liquid state can be viewed and discussed. He proposed that a liquid should be viewed as an essentially elastic medium, in which elasticity is ``masked'' by fluidity at
times larger than $\tau$ (see Ref. \cite{frenkel}).

In this approach, Frenkel proposed to consider its vibrational states using the following continuity
argument. The solid vibrational energy is calculated for an isotropic solid, glass, and is expressed in terms of normal modes  \cite{landau}. Frenkel noted that in his picture of liquid flow, no qualitative distinction can be made between a glass and a liquid: a solid glass is different from a liquid by the value of $\tau$ only (see Ref. \cite{frenkel}). Because the transition between a liquid and a glass is continuous, vibrational states of a liquid just above the glass transition temperature should not substantially change from those of a glass \cite{frenkel}. This argument points to the continuity of vibrational states between a solid and a liquid, and offers an important insight into a liquid from the solid phase.

\section{Relaxation time and vibrational states in a liquid}

At the phenomenological level, two ways to incorporate $\tau$ into the existing theories have been proposed. The first one starts from a solid phase, and modifies elasticity equations. This gives rise to frequency-dependent elastic moduli, which are interpolated between zero at $\omega\tau\ll 1$ and their solid-state value at $\omega\tau\gg 1$ \cite{frenkel}. Another way is to start from a liquid phase, and to modify the hydrodynamic equations, by either directly introducing Maxwell viscoelastic term in the Navier-Stokes equation \cite{frenkel}, or by introducing the exponential decay, with $\tau$ as decay time, of the memory function of current correlations \cite{yip}.

As reviewed recently, the microscopic understanding of liquid dynamics is still lacking \cite{rec-review}. Because elasticity and hydrodynamics describe a system at length scale that is large compared to interatomic separations, these approaches do not address the microscopic effects of relaxation: elementary relaxation processes in a liquid are local atomic jumps \cite{dyre}. Other disadvantages of the phenomenological description have been discussed in Ref. \cite{rec-review}. An open question remains as to how can the liquid ability to sustain (lose) transverse modes with frequency $\omega>1/\tau$ ($\omega<1/\tau$) be explained at the atomistic level, similar to the normal-mode analysis in a solid? We propose that one way to understand the absence of propagating shear modes with frequency $\omega<1/\tau$ in a liquid at the atomistic level is to note that local atomic rearrangement processes effectively break some of the interactions between neighbouring atoms and stop wave propagation.

A local structural rearrangement involves a jump of an atom from its cage, with subsequent relaxation of the local structure \cite{dyre}. We call these rearrangements local relaxation events (LREs). Lets consider how periodic transverse motion of atom $i$ in a liquid affects the neighbouring atom $j$. Lets assume that atom $i$ is vibrated with period $\tau_0$ in the direction perpendicular to the line connecting $i$ and $j$, exciting a shear wave, and consider atom $j$ is involved in a LRE with a period $\tau$. If atom $i$ vibrates with period $\tau_0\gg\tau$, its displacement does not affect atom $j$, because during time $\tau_0$ atom $j$ has {\it independently} jumped many times due to thermally-excited LRE. In this case, the wave does not propagate. If, on the other hand, $\tau_0\ll\tau$, $i$ has enough time to interact with atom $j$ between two consecutive LREs at site $j$ (atom $j$ is frozen on the time scale of atom $i$). In this case, the displacement of atom $i$ affects atom $j$ in the same way as in a solid, and the wave propagates. Because the vibration of atom $i$ is the superposition of modes with different frequencies, we find that modes with frequency $\omega\gg 1/\tau$ ($\omega\ll 1/\tau$) can (can not) propagate in a liquid.

We note that this discussion applies to volume-conserving shear waves. Low-frequency longitudinal (hydrodynamic) waves propagate in the presence of LREs as well, which is related to the existence of the finite zero-frequency bulk modulus in liquids (zero-frequency shear modulus in liquids is zero by definition). We also note that both high-frequency longitudinal and transverse waves are subject to some damping, but the damping of shear waves with frequency $\omega<1/\tau$ is so large that it is impossible to talk about their propagation \cite{frenkel}. Damping decreases with $\omega\tau$ in the $\omega\tau>1$ regime for longitudinal and shear waves, and increases with $\omega\tau$ in the $\omega\tau<1$ regime for longitudinal waves \cite{frenkel}.

In this picture, as we approach a liquid from the solid phase, LREs modify the solid vibrational spectrum by effectively removing the restoring force (or force constants) for shear modes at frequency $\omega<1/\tau$. As a result, the oscillatory motion starts to ``slip'' at low frequency, giving rise to diffusion due to LREs. On temperature increase, the emergent diffusive motion due to LREs contributes to the increased kinetic energy of the system $K=3NT/2$. On the other hand, we can assume that the contribution of the emergent diffusive motion to the potential energy is small. Lets increase the system temperature $T$ by a small amount $\Delta T$, corresponding to a certain new value of $\tau$. As restoring forces for shear modes with frequency $\omega<1/\tau$ are removed, atoms start slipping and diffusing along the same directions at which they oscillated in the presence of restoring forces at temperature $T$ (albeit with larger displacements). Note that interatomic interactions that give rise to the restoring forces at temperature $T$ and interactions experienced by the newly diffusing atoms at temperature $T+\Delta T$ have the same microscopic origin. Therefore, the absence of restoring forces at frequency $\omega<1/\tau$ means that the emergent diffusive motion weakly contributes  to the system's potential energy: a large value of this contribution would imply strong interactions at frequency $\omega<1/\tau$ and, therefore, the existence of restoring forces and shear modes with this frequency. Hence, the contribution of the potential energy of the emergent diffusive motion to the total energy is small compared to other energy terms (the potential energy of the unmodified longitudinal mode and the kinetic energy $K$), and can be ignored.


\section{Liquid energy}

The glass transition temperature $T_g$ is defined from the condition $\tau(T_g)=10^3$ s \cite{dyre}. Above $T_g$, each atom participates in the vibrational motion during time $\tau$ and in the diffusional motion when it jumps between two equilibrium positions. The instantaneous liquid energy, $E_{ins}$, is the sum of the vibrational, $E_{vib}$, and the diffusional, $E_{dif}$, components:

\begin{equation}
E_{ins}=E_{vib}+E_{dif}
\end{equation}

\noindent Within thermal fluctuations, $E_{ins}$ is equal to the liquid energy averaged over time $t\gg\tau$.

Lets now approach the liquid from the solid phase as discussed in the previous section. In the solid glass phase at $T\le T_g$, $E_{dif}=0$, and the energy is purely vibrational and consists of kinetic and potential energy of the oscillatory motion. Lets assume that the temperature is instantaneously raised, corresponding to a certain value of $\tau<\tau(T_g)$ in the equilibrium state. As discussed in the previous section, the effect of local rearrangements processes is to effectively remove the potential energy component that give rise to the restoring force for two transverse modes with frequency $\omega<1/\tau$. The energy of such instantaneously prepared state, $E$, is the sum of the kinetic energy, $K$, the potential energy of one longitudinal mode, $U_l$, and the potential energy of two shear modes of frequency $\omega>1/\tau$, $U_t(\omega>1/\tau)$:

\begin{equation}
E=K+U_l+U_t(\omega>1/\tau)
\end{equation}

We now let the system evolve until the equilibrium state is established. In the equilibrium state, $E$ becomes the energy of vibrating and diffusing atoms. From the energy conservation, the energies of the initial and the final state are the same: $E=E_{ins}$. Hence, we can use Eq. (2) to calculate the liquid energy.

Because $K=K_l+K_t$, where $K_l$ and $K_t$ are the kinetic energies of longitudinal and transverse modes, $E=E_l+K_t+U_t(\omega>1/\tau)$, where $E_l$ is the energy of the longitudinal mode. If $E_t$ is the energy of the transverse mode, $K_t=U_t=\frac{E_t}{2}$ from the equipartition theorem. Because the equipartition theorem applies to the energy of a single oscillator as well as to the sum over any set of oscillators, $U_t(\omega>1/\tau)=\frac{E_t(\omega>1/\tau)}{2}$. Then, $E=E_l+\frac{E_t}{2}+\frac{E_t(\omega>1/\tau)}{2}$. $E_t$ can be written as the sum over modes below and above $\omega=1/\tau$, $E_t=E_t(\omega<1/\tau)+E_t(\omega>1/\tau)$, where the two terms refer to their solid-state values. Then,

\begin{equation}
E=E_l+E_t(\omega>1/\tau)+\frac{E_t(\omega<1/\tau)}{2}
\end{equation}

The first two terms in Eq. (3) are the vibrational energies of one longitudinal mode and two transverse modes with frequency $\omega>1/\tau$. In the solid glass phase, where $\tau$ is very large, these two terms become the vibrational energy of the system, and the last term is zero. On increasing the temperature in a liquid, the second term decreases. The last term ensures that the total kinetic energy of the system does not change as a result of this decrease: the kinetic energy is defined by temperature only ($K=3NT/2$) regardless of how it partitions into vibrational and diffusional motion.

To calculate the first two terms in Eq. (3), we separate $N$ longitudinal and $2N$ transverse normal modes with frequency $\omega>1/\tau$ in the statistical sum of the vibrational motion:

\begin{eqnarray}
Z=h^{-3N}\int\exp\left(-\frac{1}{2T}\sum\limits_{i=1}^N(p_i^2+\omega_{li}^2q_i^2)\right) dpdq\\\times
\int\exp\left(-\frac{1}{2T}\sum\limits_{\omega_{ti}>\omega_0}^{2N}(p_i^2+\omega_{ti}^2q_i^2)\right)dpdq\nonumber
\label{part}
\end{eqnarray}

\noindent where $\omega_{li}$ and $\omega_{ti}$ are frequencies of longitudinal and transverse vibrations and $\omega_0=1/\tau$. This gives $Z=Z_l\cdot Z_t(\omega>1/\tau)=(2\pi T)^N\left(\prod\limits_{i=1}^N \omega_{li}\right)^{-1}\times(2\pi T)^{N_1}\left(\prod\limits_{\omega_{ti}>\omega_0}^{2N} \omega_{ti}\right)^{-1}$, where $Z_l$ and $Z_t(\omega>1/\tau)$ are the contributions to $Z$ due to the longitudinal mode and the transverse modes with frequency $\omega>1/\tau$, respectively, and $N_1$ is the number of transverse modes with $\omega>\omega_0$. The corresponding energies are $E_l=T^2\frac{d}{dT}\ln(Z_l)=NT$ and $E_t(\omega>1/\tau)=T^2\frac{d}{dT}\ln(Z_t(\omega>1/\tau))=N_1T$. This gives the first two terms in Eq. (3).

$N_1$ can be calculated from the density of states in the Debye model. Recall that in an isotropic solid (glass), vibrational density of states is written as $g(\omega)=\frac{V}{2\pi^2}\left(\frac{1}{c_l^3}+\frac{2}{c_t^3}\right)\omega^2$, where $c_l$ and $c_t$ are longitudinal and transverse sound velocities \cite{landau}. The Debye frequency $\omega_m$ is defined from the condition $\int\limits_0^{\omega_m} g(\omega){\rm d}{\omega}=3N$, where $N$ is the number of atoms in the system. If the mean speed of sound is introduced as $\frac{3}{\bar{c}^3}=\frac{2}{c_t^3}+\frac{1}{c_l^3}$, the density of states becomes $g(\omega)=\frac{9N\omega^2}{\omega_m^3}$, where $\omega_m^3=6\pi^2N\bar{c}^3/V$ \cite{landau}. Instead of $\omega_m$, lets define transverse Debye frequency, $\omega_{mt}$, using the condition $\int\limits_0^{\omega_{mt}}g_t(\omega){\rm d}{\omega}=2N$, which reflects the fact that $2N$ degrees of freedom are associated with transverse vibrations. Here, $g_t(\omega)=\frac{V}{2\pi^2}\frac{2}{c_t^3}\omega^2$ is the density of states of transverse vibrations. This gives $\omega_{mt}^3=6\pi^2Nc_t^3/V$ and $g_t(\omega)=\frac{6N}{\omega_{mt}^3}\omega^2$. Then, $N_1=\int\limits_{\omega_0}^{\omega_{mt}}g_t(\omega)d\omega=2N\left(1-\left(\frac{\omega_0}{\omega_{mt}}\right)^3\right)$, and the vibrational energy of one longitudinal and two transverse modes with frequency $\omega>1/\tau$ becomes $E_{vib}=E_l+E_t(\omega>1/\tau)=(N+N_1)T=NT\left(3-2\left(\frac{\omega_0}{\omega_{mt}}\right)^3\right)$.

To calculate the last term in Eq. (3), we note that similarly to $E_t(\omega>1/\tau)=N_1 T$, $E_t(\omega<1/\tau)=N_2 T$, where $N_2$ is the number of transverse modes with $\omega<\omega_0$. Because $N_2=2N-N_1$, $N_2=2N\left(\frac{\omega_0}{\omega_{mt}}\right)^3$. Then, $E=(N+N_1+N_2/2)T$, giving

\begin{equation}
E=NT\left(3-\left(\frac{\omega_0}{\omega_{mt}}\right)^3\right)
\label{8}
\end{equation}

Eq. (\ref{8}) predicts the relationship between the energy of a liquid and its relaxation time $\tau=1/\omega_0$. The first term in Eq. (\ref{8}) is the classical energy of a solid. The second term is the contribution to the energy due to local rearrangement processes in a liquid. According to Eq. (\ref{8}), liquid energy is equal to the solid-state value of $3NT$ in the very broad range of relaxation time $\tau$. This range starts from the experimentally accessible $\tau=10^3$ s, and continues up to the Debye vibrational period of $\tau_{\rm D}=10^{-13}$ s. Only when $\omega_0$ becomes of the order of Debye frequency, liquid energy and heat capacity start decreasing from their solid-state values. Hence, Eq. (\ref{8}) predicts that this decrease should start in low-viscous liquids not far above their melting point. This agrees with experiments on liquid Hg, K, Na and other systems \cite{grimvall,wallace}. If $\omega_0=\omega_{mt}\approx1/\tau_{\rm D}$, Eq. (\ref{8}) gives the lower limit of the energy of a simple monatomic liquid of $2NT$. This is the result that Brillouin obtained under the assumption of the complete loss of transverse modes \cite{bril}.

The reason why the energy of a liquid is equal to the solid-state value of $3NT$ unless $\omega_0$ is comparable to the Debye frequency can also be discussed on the basis of Eq. (1). At any given moment of time, the number of particles in the transient diffusional state, $N_{dif}$, is $N_{dif}=N_0\exp(-V/T)$, where $V$ is an activation barrier for a LRE ($V$ can be temperature-dependent) and $N_0$ is the total number of atoms. Because $\tau=\tau_{\rm D}\exp(V/T)$ (see Ref. \cite{frenkel}), $\frac{N_{dif}}{N_0}=\frac{\tau_{\rm D}}{\tau}$. When $\tau\gg\tau_{\rm D}$, $N_{dif}/N_0\ll 1$, i.e. most of the particles in the system are vibrating, and only a negligibly small proportion of particles are diffusing. Because $N_{dif}/N_0\ll 1$, the contribution to the energy from diffusion can be ignored in Eq. (1), giving $E_{ins}=E_{vib}$. As we have seen above, the vibrational energy, $E_{vib}=(N+N_1)T=3NT$ when $\tau\gg\tau_{\rm D}$, i.e. the same as in the solid. This point will be discussed below for its implications for the problem of glass transition.

\section{Heat capacity}

From Eq. (\ref{8}), $c_v=\frac{1}{N}\frac{{\rm d}E}{{\rm d} T}$ is:

\begin{equation}
c_v=\frac{\rm d}{{\rm d} T}\left( T\left(3-\left(\frac{\tau_{mt}G_{\infty}}{\eta}\right)^3\right)\right)
\label{10}
\end{equation}

\noindent where $\tau_{mt}$ is the Debye period for transverse vibrations, $\tau_{mt}\approx\tau_{\rm D}$, $\eta$ is viscosity, and we used the Maxwell relation $\eta=\tau G_{\infty}$ ($G_{\infty}$ is instantaneous shear modulus) in order to directly relate $c_v$ and viscosity. Note that Eq. (\ref{10}) relates heat capacity and viscosity with no fitting parameters, because $\tau_{mt}G_{\infty}$ is given by the liquid properties.

Mercury is an appropriate liquid to test the prediction of Eq. (\ref{10}) for a number of reasons: it is a simple monatomic liquid, anharmonic effects in mercury are small \cite{wallace} (note that thermal expansion is ignored in our discussion), and finally, it is a low-viscous liquid, which makes it possible to observe the decrease of heat capacity in the reasonable temperature range: as discussed above, this is expected to take place when $\omega_0$ becomes of the order of Debye frequency.

We have taken measurements of $c_v$ of mercury from Ref. \cite{grimvall}, which has electronic contribution subtracted, hence giving heat capacity due to ions only. Viscosity data was taken from Ref. \cite{carlson}, and interpolated in order to use in Eq. (\ref{10}) to calculate $c_v$. Because viscosity shows the deviation from the Arrhenius behaviour, we used the Vogel-Fulcher-Tammann expression to fit the data, $\eta=\eta_0\exp(A/(T-T_0))$. A good fit was achieved with $\eta_0$=0.612$\cdot10^{-3}$ Pa$\cdot$s, $A$=235.157 K and $T_0$=40.726 K (see Fig. 1a). Because the experimental temperature range of viscosity is 250--600 K \cite{carlson}, slightly smaller than the experimental temperature range of $c_v$, 234--754 K \cite{grimvall}, we used the fitted expression to extrapolate viscosity data in the temperature range of $c_v$ (see Fig. 1).

\begin{figure}
\begin{center}
{\scalebox{0.55}{\includegraphics{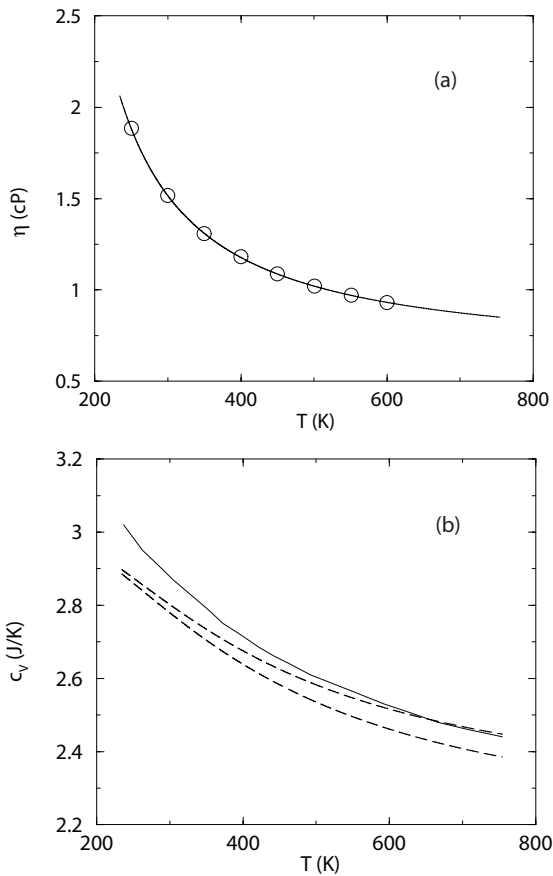}}}
\end{center}
\caption{(a) Circles are experimental data of viscosity of mercury, line is the fit and extrapolation using the Vogel-Fulcher-Tammann formula; (b) Solid line is the experimental $c_v$ for mercury. Dashed lines are calculated values, using $\tau_{mt}G_{\infty}$ of 0.55 $\cdot$10$^{-3}$ Pa$\cdot$s (top) and 0.57 $\cdot$10$^{-3}$ Pa$\cdot$s (bottom).}
\end{figure}

Eq. (\ref{10}) has no fitting parameters, hence there is no flexibility to fit its prediction to the experimental curve of $c_v$. However, because $\tau_{mt}$ and $G_{\infty}$ are not known precisely for mercury, in practice we can use $\tau_{mt}G_{\infty}$ as a fitting parameter. In Fig. 1b, we compare $c_v$, calculated from Eq. (\ref{10}), with the experimental data, using $\tau_{mt}G_{\infty}$=0.55$\cdot$10$^{-3}$ and 0.57$\cdot$10$^{-3}$ Pa$\cdot$s. Given that the experimental error of $c_v$ is 0.1--0.2 J/K \cite{grimvall} (experimental error of viscosity is not known), we find a good agreement between the calculated and experimental $c_v$.


It is important to note that the used value of $\tau_{mt}G_{\infty}$ in Fig. 1b gives physically sensible values for $\tau_{mt}$ and $G_{\infty}$: if $G_{\infty}$ of mercury is about 5 GPa \cite{wallace1}, the used values of $\tau_{mt}G_{\infty}$ imply $\tau_{mt}$ of about 0.1 ps, a typical value of Debye vibrational period. We also note that the used value of $\tau_{mt}G_{\infty}$ is close to that in other liquids: in sodium, for example, the product of high-temperature $\tau$ and $G_{\infty}$ is about $0.5\cdot10^{-3}$ Pa$\cdot$s \cite{pilgrim}. We can therefore conclude that the decrease of $c_v$ follows, within the experimental error, the decrease of viscosity reasonably well, with physically realistic values of $\tau_{mt}$ and $G_{\infty}$.

\section{Vibrational energy of a quantum liquid}

In most liquids, crystallization or vitrification take place at low temperature. However, if a liquid state exists at low temperature, quantum effects become important. Below we calculate the effect of relaxation process on the vibrational energy of a quantum liquid, $E_{vib}$.

We first note that in considering the phonon excitations in a quantum liquid at very low temperature, Landau assumed that these are due to the longitudinal phonons only \cite{landau}. The vibrational energy is then three times smaller than that of a quantum solid, reflecting the assumption that excitations due to two transverse modes are completely lost \cite{landau}. At first glance, the discussion in the previous sections is at odds with this assumption. Indeed, we have discussed that transverse waves with frequency $\omega>1/\tau$ propagate in liquids, and their frequency range only increases at low temperature. We have seen that transverse waves disappear at high temperature only, whereas neglecting them at low temperature gives, in classical case, $C_v=2N$, inconsistent with the experiments. Interestingly, we show that transverse waves can not be excited in a quantum liquid in the low-temperature limit, confirming the Landau assumption.

In a quantum solid, $E_{vib}$ is the sum of energies of quantum oscillators, represented by one longitudinal and two transverse modes. In a quantum liquid, $E_{vib}$ is the sum of energies of one longitudinal and two transverse modes with frequency $\omega>1/\tau$. As in the Debye model of a solid, the summation over the phonon frequencies can be substituted by integration using the Debye density of states \cite{landau}. Similar to $\omega_{mt}$ introduced previously in the density of states of transverse vibrations, $g_t(\omega)=\frac{6N}{\omega_{mt}^3}\omega^2$, we define the Debye longitudinal frequency, $\omega_{ml}$, from the condition $\frac{1}{c_t^3}\frac{V}{2\pi^2}\int\limits_0^{\omega_{ml}}\omega^2{\rm d}{\omega}=N$, which reflects the fact that $N$ degrees of freedom are associated with longitudinal vibrations. This gives $\omega_{ml}^3=6\pi^2Nc_t^3/V$, and the density of states of longitudinal vibrations, $g_l(\omega)=\frac{V}{2\pi^2}\omega^2\frac{1}{c_l^3}=\frac{3N}{\omega_{ml}^3}\omega^2$. $E_{vib}$ becomes

\begin{equation}
E_{vib}=E_0+\frac{3N}{\omega_{ml}^3}\int\limits_0^{\omega_{ml}} E_\omega(T)\omega^2{\rm d}{\omega}+\frac{6N}{\omega_{mt}^3}\int\limits_{\omega_0}^{\omega_{mt}} E_\omega(T)\omega^2{\rm d}{\omega}
\label{bf}
\end{equation}

\noindent where $E_0$ is the energy of zero-point vibrations and $E_\omega(T)=\frac{\hbar\omega}{\exp\left(\frac{{\hbar\omega}}{T}\right)-1}$, giving

\begin{eqnarray}
E_{vib}=E_0+NTD\left(\frac{\hbar\omega_{ml}}{T}\right)+2NTD\left(\frac{\hbar\omega_{mt}}{T}\right)-\\
2NT\left(\frac{\omega_0}{\omega_{mt}}\right)^3D\left(\frac{\hbar\omega_0}{T}\right)\nonumber
\end{eqnarray}

\noindent where $D(x)=\frac{3}{x^3}\int\limits_0^x\frac{z^3{\rm d}z}{\exp(z)-1}$ is the Debye function \cite{landau}.

The first three terms in Eq. (8) is the vibrational energy of a quantum solid, obtained from Eq. (\ref{bf}) if $\omega_0=0$. The last term in Eq. (8) is the contribution from the relaxation process in a liquid. In the high-temperature limit, when $D(x)=1$, $E_{vib}$ becomes the vibrational energy of a classical liquid that follows from Eq. (4).

The low-temperature limit of the Debye model corresponds to the large value of the argument of $D(x)$, so that neglecting exponentially small terms, $D(x)=\frac{\pi^4}{5x^3}$ and $E_{vib}\propto T^4$ \cite{landau}. In Eq. (8), this takes place when $\frac{\hbar\omega_0}{T}\gg 1$ (which also implies $\frac{\hbar\omega_{mt}}{T}\gg 1$, because $\omega_0<\omega_{mt}$). When $D(x)\propto\frac{1}{x^3}$, the last two terms in Eq. (8) disappear for any $\omega_0$, corresponding to the complete loss of excitations due to transverse waves. This can be understood because at temperature $T$, only phonons with frequency $\hbar\omega\le T$ are excited and contribute to $E_\omega$ and the temperature-dependent energy terms in Eq. (\ref{bf}). Hence, in the low-temperature limit $\hbar\omega_0\gg T$ only phonons with frequency $\omega\ll\omega_0$ are excited, but transverse phonons with this frequency are excluded from Eq. (\ref{bf}). As a result, transverse modes can not be excited in a liquid in the low-temperature limit, and all excitations in the system are due to the longitudinal phonons only.

\section{Implications for glass transition}

There are two implications of the discussed picture for the long-standing problem of glass transition \cite{dyre,angell,langer}. The first one concerns the general approach to glass transition from a liquid state. Basic discussions of liquids begin with stochastic dynamics, which describes Brownian motion and self-diffusion, and is modeled by Langevin-type and related equations \cite{yip,zwanzig,hansen}. These approaches assume stochastic dynamics, corresponding to the regime $\tau\approx\tau_{\rm D}$. To discuss glass transition, the formalism is then modified so that is can be extended to the supercooled state where $\tau\gg\tau_{\rm D}$ and eventually to the glass phase where $\tau(T_g)=10^3$ s. The important point here is that the condition $\tau\gg\tau_{\rm D}$ holds in 15 decades of $\tau$ ($10^{-12}$--10$^3$ s). In this range of $\tau$, the system is a liquid as follows from its inability to support static shear stress on experimental time scale, i.e. it flows, yet its main properties are those of a solid. First, as discussed above, the liquid energy is equal to the vibrational energy of a solid in the range of 15 decades of $\tau$. Second, a liquid supports solid-like shear modes with frequencies $\omega>1/\tau$ which can also span up to 15 orders of magnitude depending on $\tau$. Hence, in almost entire glass transformation range, the important properties of a liquid are governed by its solid-like elastic features. This suggests that a glass transition theory can include the solid-like elastic properties of a liquid from the outset, rather than modify and extend the stochastic high-temperature behaviour (where these properties are absent) to the supercooled liquid state and glass. Basing on this view, we have recently proposed how glass transition can be understood on the basis of liquid elasticity \cite{tau}.

The second implication concerns the behaviour of liquid heat capacity at $T_g$. Opposite to the increase of $c_v$ on lowering the temperature in Fig. 1, constant-pressure heat capacity decreases at $T_g$, often with a jump, $\Delta c_p$  \cite{dyre,angell,langer}. It is therefore important to discuss how the two effects can be reconciled. Several popular theories of glass transition relate $\Delta c_p$ at $T_g$ to the reduction of liquid configurational entropy and to an underlying phase transition \cite{dyre}. This approach has been convincingly criticized for a number of reasons \cite{dyre}. We propose that $\Delta c_p$ at $T_g$ can be understood as a kinetic effect related to the non-equilibrium nature of glass transition, rather than to the presence of an underlying phase transition. Hence, this effect is different from the one that governs liquid heat capacity in Eq. (6): recall that we related liquid heat capacity to the liquid equilibrium vibrational states.

$\Delta c_p$ at $T_g$ can be understood from the disappearance of LREs at $T_g$. The presence of these processes in a liquid (or their absence in a glass) has a profound effect on several of important properties of the system. First, LREs are accompanied by the increases of local volume. This is because the energy needed for an atom to escape its cage at the constant volume is very large due to strong short-range interatomic repulsions. Hence, atoms in the cage need to increase its volume in order to allow for the escape of the central atom \cite{frenkel,dyre}. Because $T_g$ is defined from the condition of disappearance of LREs on the experimental time scale, thermal expansion is larger (smaller) above (below) $T_g$. This gives rise to the experimentally observed kink of density at $T_g$ \cite{doremus}, and contributes to the jump of $c_p$ at $T_g$. Second, because LREs are accompanied by fluctuating local volume increases, they result in a faster softening of the high-frequency elastic constants of a liquid as compared to a glass, as witnessed by the kink of the high-frequency longitudinal and transverse sound velocities \cite{solt,scarponi}. This introduces a difference in the compressibility between a liquid and a glass at $T_g$, and contributes to $\Delta c_p$. Finally, at times $t\gg\tau$, LREs ensure that the configurational entropy of the system, $s_c$, is equal to the ``entropy of melting'' associated with the increase of the number of configurations of a liquid over that of a solid \cite{ziman}. If LREs are frozen at $T_g$, the change of $s_c$ contributes to $\Delta c_p$.

Importantly, small variations of temperature around $T_g$ give large changes of $\tau$, leading to the freezing of LREs in a very narrow temperature range and therefore giving rise to the observed apparent jump of $c_p$. Generally, this is because at $T_g$, the activation barrier for LREs, $V$, is very large: because $\tau=\tau_{\rm D}\exp(V/T)$ and $\tau(T_g)=$100-1000 s, $V/T_g=$35-37. Moreover, in ``fragile'' liquids \cite{angell}, i.e. liquids with large slope of $\ln(\tau)$ vs $T_g/T$ at $T_g$, small variations of temperature around $T_g$, $\Delta T$, result in a large change of $\tau$. Take, for example, propylene carbonate, for which relaxation time follows the VFT law, $\tau=\tau_{\rm D}\exp(A/(T-T_0))$ with parameters $A=612$ K, $T_0=139$ K and $\log(\tau_{\rm D})=-12.3$ s \cite{casa}. The conditions $\tau(T_g)=100$ s and $\tau(T_g-\Delta T)=1000$ s give $T_g\approx 157.6$ K and $\Delta T\approx 1.2$ K, i.e. the decrease of temperature at $T_g$ by only 1.2 K gives a ten-fold increase of $\tau$. Similarly, for several other fragile liquids \cite{casa}, we find that the variations of $\Delta T=\pm$(1-2) K at $T_g$ result in the variations of $\tau$ by an order of magnitude.

This picture predicts that no $\Delta c_p$ would have been observed had the experiment lasted for time $t\gg\tau$. This is consistent with the general finding that the observed anomalies at $T_g$ depend on cooling rate and observation time. Another prediction of this picture is that $\Delta c_p$ at $T_g$ should be sharp in fragile, but not in ``strong'' liquids, which have a smaller slope of $\ln(\tau)$ vs $T_g/T$ at $T_g$. This is consistent with the experimental results \cite{angell}. To summarize, $\Delta c_p$ can be understood as a natural signature of the glass transition insofar as this transition is defined by the freezing of LREs at the experimental time scale, but not as a result of a phase transition.

\section{Conclusions}

In summary, we discussed the solid-like elastic properties of liquids, including its vibrational states, energy and heat capacity. The decrease of liquid heat capacity with temperature was attributed to the increasing loss of two transverse modes with frequency $\omega<1/\tau$. In a simple model, we related liquid heat capacity and viscosity, and compared this relation with the experimental data of mercury. We also calculated the vibrational energy of a quantum liquid, and showed that transverse phonons can not be excited in the low-temperature limit. Finally, we discussed the implications of the proposed approach to liquids for the problem of glass transition.

I am grateful to Prof. V. V. Brazhkin, E. Artacho, M. T. Dove and J. C. Phillips for discussions and to EPSRC for support.

\end{document}